\documentclass{article}

\usepackage[main, final, nonatbib]{neurips_2026}
\usepackage{amsmath}
\usepackage{graphicx}

\usepackage[utf8]{inputenc}
\usepackage[T1]{fontenc}
\usepackage{hyperref}
\usepackage{url}
\usepackage{booktabs}
\usepackage{amsfonts}
\usepackage{nicefrac}
\usepackage{microtype}
\usepackage{xcolor}
\usepackage{siunitx}
\usepackage[style=numeric,backend=biber]{biblatex}
\title{Symplectic Normal Coordinates Flow for Generative Modeling of Hamiltonian Systems}

\author{%
  Christopher M.~Pierce \\
  Physical Sciences Division\\
  University of Chicago\\
  Chicago, IL 60637 \\
  \texttt{cmpierce@uchicago.edu} \\
}

\begin{document}

\maketitle

\begin{abstract}
In conditioned generative models of physical systems, the densities at different values of the conditioning parameters are often related through symplectic transformations.
For stable linear Hamiltonian systems, there exists a symplectic transformation to ``normal coordinates'' such that the dynamics is reduced to rotations in each phase-space plane.
This picture extends to the nonlinear case where, away from resonances, the Birkhoff normal form again consists of rotations in phase space, now with amplitude dependence.
In this work, we introduce an invertible symplectic flow layer consisting of amplitude-dependent rotations in learned normal coordinates.
We derive the condition under which the nonlinear rotation layer is symplectic and detail an architecture that guarantees it.
Experiments are carried out on three conditional generative modeling problems: a synthetic chain of coupled nonlinear oscillators and two problems from accelerator physics.
We show empirically that normalizing flows with the new layers achieve competitive or superior negative log-likelihoods compared to baseline flows, with gains that grow with the strength of the nonlinearity on the synthetic system.
\end{abstract}

\section{Introduction}

Discovery of the model architectures that naturally express the symmetries of problem domains has been a fruitful line of research in generative machine learning~\cite{bronsteinGeometricDeepLearning2021,hoogeboomEquivariantDiffusionMolecule,kohlerEquivariantFlowsExact}.
For datasets that are rooted in classical mechanics, symplectic geometry is the language which describes their evolution in time.
We are thus motivated to respect its symmetries when building generative models that represent data drawn from distributions that have evolved in time according to the rules of classical mechanics.
This is particularly important for models that are conditioned on time or on system parameters which are related by symplectic transformation (e.g., the two conditional distributions with different system parameters can be evolved backwards in time to a common initial state).

There is a growing body of research on generative models that possess the symmetries of symplectic geometry.
However, it is not enough to just have models that possess the symmetries of the data.
It is also important to find the ones that are biased towards expressing the typical evolution of the types of systems being modeled.
As an example, it is well known that multilayer perceptrons (MLPs) possess spectral bias causing them to learn low frequency information first~\cite{basriFrequencyBiasNeural,rahamanSpectralBiasNeural}.
Architectural enhancements such as Fourier features~\cite{tancik_fourier_2020} exist which overcome these biases and are necessary for learning data with high frequency content.
For models of physical systems, we are interested in models that are biased towards ``typical'' time evolution.

In this work, we introduce a new symplectic flow based on normal coordinates and inspired by Birkhoff normal form, and show through experiments that it is well suited for learning data from time-dependent weakly nonlinear Hamiltonian systems.
We open-source the normal coordinates flow for the purpose of open science and building on top of it.\footnote{Code is available at \url{https://github.com/electronsandstuff/Normal-Coordinates-Flow}.}
The remainder of this article is organized as follows.
In Sec.~\ref{sec:related-work}, we discuss the related research in this area.
We then give the physics and mathematics background necessary to understand the rest of the work in Sec.~\ref{sec:background}.
The new architecture is described in Sec.~\ref{sec:architecture}, and we show how it can be integrated into a normalizing flow for generative modeling.
In Sec.~\ref{sec:experiments}, experiments using the flow to model a synthetic oscillator chain and data from particle accelerators are presented.
Finally, we discuss limitations and conclude in Sec.~\ref{sec:conclusion}.

Our core contributions are the following.
\begin{itemize}
	\item A conditioned flow layer: the normal coordinates flow. It consists of a symplectic transformation to normal coordinates, an amplitude-dependent rotation layer, and a symplectic transformation out of normal coordinates.
	\item Derivation of the conditions under which the nonlinear rotation layer remains symplectic and demonstration that they are met by a GradNet~\cite{chaudhariGradientNetworks2025}.
	\item An integrated normalizing flow model consisting of NVP layers~\cite{dinhDensityEstimationUsing2017} followed by normal coordinates flows capable of learning conditioned data parameterized in a symplectic manner.
	\item Empirical results on weakly nonlinear Hamiltonian systems showing good performance of the normal coordinates flow model compared to baseline models.
	\item Ablation studies on the normal coordinates flow to quantify the importance of choices within the architecture.
\end{itemize}

\section{Related Work}
\label{sec:related-work}

\paragraph{Symplectic Flows}

Our work builds on previous flow layers that respect the symmetries of symplectic geometry.
One approach involves integrating the equations of motion from a learned Hamiltonian for the system~\cite{greydanusHamiltonianNeuralNetworks,aichSymplecticGenerativeNetworks2025,holderriethHamiltonianScoreMatching,rezendeEquivariantHamiltonianFlows2019}.
Generating functions are also a natural starting point for making parameterized canonical transformations~\cite{chenDatadrivenPredictionGeneral}.
These require derivatives to be taken with respect to the phase space coordinates to calculate the transformation and may not be invertible, making them difficult to use in normalizing flows.
Some methods build up the symplectic map from subclasses of general canonical transformations which are parameterizable~\cite{liNeuralCanonicalTransformation2020}.
Some techniques avoid the complications of taking derivatives or integrating equations of motion by finding flow layers that are symplectic and can approximate all transformations.
SympNets, for example, have had different forms consisting of linear transformations and nonlinear activation layers as well as gradient networks~\cite{jinSympNetsIntrinsicStructurepreserving2020a,jinOptimalUnitTriangular2022,xiongNonseparableSymplecticNeural2022}.
HenonNet uses a similar approach of smaller symplectic flow layers, but uses the gradient of a potential function to produce a nonlinear mapping~\cite{burbyFastNeuralPoincare2021}.

\paragraph{Normal Coordinates and Action-Angle Coordinates Flows}

The closest related work to the flow explored in this article is the use of a transformation to normal coordinates for surrogate modeling of nearly-periodic symplectic maps.
In Ref.~\cite{duruisseauxApproximationNearlyperiodicSymplectic2023}, promising results are shown for a structure preserving network capable of learning a fixed-time-step map for the evolution of autonomous systems.
The work does not explore generative modeling, conditioning, time dependence of the system, or action-dependent rotations.
Another related architecture is the ``action-angle network,'' where a learned map transforms the data from integrable systems into action-angle coordinates.
Here, a learnable dynamics model which depends on the actions is used to evolve the angles for some time step~\cite{daigavaneLearningIntegrableDynamics2022}.
Conservation of the action over the full trajectory is enforced through soft constraints.
Learning transformations to action-angle coordinates has also been used to reveal the symmetries in integrable systems~\cite{bondesanLearningSymmetriesClassical2019}.
Existing work has focused on learning symplectic maps for autonomous systems.
We extend this line of research to generative modeling of time-dependent systems with conditioned amplitude-dependent evolution of the samples in normal coordinates without a numerical ODE solver.

\section{Background}
\label{sec:background}

In this section, a concise overview is given of symplectic geometry and its connection to classical mechanics.
We then describe normal coordinates and related concepts for nonlinear systems, which are the inspiration for this architecture.

\paragraph{Symplectic Geometry}
A symplectic manifold $(M, \omega)$ is a differentiable manifold $M$ equipped with a closed, nondegenerate 2-form $\omega$ (at each point, an antisymmetric object which takes two vectors from the tangent space and outputs a real number).
These manifolds are always of even dimension, and by Darboux's theorem, coordinates can be found and partitioned in a standard way $(q_1, \ldots, q_n, p_1, \ldots, p_n)$ such that the symplectic form is locally~\cite{holmWHATSymplecticGeometry2016},
\[\omega = \sum_{i=1}^n \mathrm{d}q_i\wedge \mathrm{d}p_i, \qquad \omega(v, w) = v^T \Omega w, \qquad \Omega = \begin{pmatrix} 0 & I_n \\ -I_n & 0 \end{pmatrix}.\]
Here, we show the evaluation of the 2-form on a set of tangent vectors in matrix notation in the second expression, and the reader should note that the matrix $\Omega$ is reminiscent of complex algebra ($\Omega^2 = -I$).
In $\mathbb{R}^{2n}$, where the rest of this work will live, this description suffices globally.
A transformation $T: M \to M$ is symplectic if it preserves the symplectic form everywhere: $T^*\omega = \omega$.
This is equivalent to requiring $J^T\Omega J=\Omega,$ where $J$ is the Jacobian of $T$ in Darboux coordinates and the condition holds at every point on the manifold.

\paragraph{Hamiltonian Mechanics}
In Hamiltonian mechanics, the configuration of the physical system is described by a point $z = (q_1, \ldots, q_n, p_1, \ldots, p_n)$ in phase space and its dynamics is described by the real-valued function $H(z, t)$ defined everywhere in phase space, called the Hamiltonian.
Motivated by our discussion of symplectic geometry, we write the equations of motion as,
\[\dot{z} = \Omega\nabla_z H(z, t),\]
where $\nabla_z H(z)$ is the gradient of the Hamiltonian with respect to the coordinates partitioned as in our definition of $z$, and we take its product with the matrix $\Omega$ from earlier.
In this form, it is now clear that symplectic transformations are exactly those which preserve the form of the equations of motion in Hamiltonian mechanics, such as the transformations representing time evolution of the system.

\paragraph{Normal Coordinates}
Motivated by applications to particle accelerators in the 1950s, the time dependent harmonic oscillator / the Hill equation ($H(p, q, s) = p^2/2 + K(s)q^2/2$) was investigated with solutions of the form $q(s) = \sqrt{\varepsilon\beta(s)}\cos(\phi(s) + \phi_0)$, $p(s)=q^\prime(s)$, where $s$ is a time-like variable and $\varepsilon$ is the Courant--Snyder invariant which factors out of the equations of motion~\cite{courantTheoryAlternatinggradientSynchrotron1958}.
The function $\beta(s)$ solves the equations of motion if it satisfies a nonlinear envelope equation and the phase is computed from it by integration as $\phi(s) = \int \mathrm{d}s/\beta(s)$.
For periodic systems, $\beta(s)$ is uniquely determined by $K(s)$, where for initial value problems we also require knowledge of $\beta(s_0)$ and $\alpha(s_0) = -\beta^\prime(s_0)/2$. The parameters $(\beta(s), \alpha(s))$ at every time define a canonical transformation to ``normal coordinates'' such that evolution in phase space is simply a rotation by the time-dependent angle $\phi(s)$.
That is, the transport matrix $M(s|s_0)$ which transforms $z = (q, p)$ as $z(s) = M(s|s_0)z(s_0)$ can be decomposed as,
\[M(s|s_0) = N(s)^{-1}R(s, s_0)N(s_0),\]
where the ``normalizing transformation'' $N(s)$ and rotation matrix $R(s, s_0)$ can be written out explicitly as,
\[N(s) = \begin{pmatrix} 1/\sqrt{\beta(s)} & 0 \\ \alpha(s)/\sqrt{\beta(s)} & \sqrt{\beta(s)} \end{pmatrix}, \qquad R(s, s_0) = R(\Delta\phi) = \begin{pmatrix} \cos(\Delta\phi) & \sin(\Delta\phi) \\ -\sin(\Delta\phi) & \cos(\Delta\phi) \end{pmatrix},\]
where $\Delta \phi = \phi(s) - \phi(s_0)$.
For stable linear periodic Hamiltonian systems in higher dimensions, we can still perform a symplectic transformation to a series of normal modes in each conjugate variable ($q_i$ and $p_i$)~\cite{qinNecessarySufficientCondition2019,qinGeneralizedCourantSnyderTheory2013,lebedevBetatronMotionCoupling2010}.
Here, each phase space plane evolves as,
\[\zeta(s) = \begin{pmatrix} C(\Delta\phi) & S(\Delta\phi) \\ -S(\Delta\phi) & C(\Delta\phi) \end{pmatrix} \zeta(s_0), \qquad \begin{aligned} C(\Delta\phi) &= \mathrm{diag}\big(\cos\Delta\phi_1, \ldots, \cos\Delta\phi_n\big), \\ S(\Delta\phi) &= \mathrm{diag}\big(\sin\Delta\phi_1, \ldots, \sin\Delta\phi_n\big), \end{aligned}\]
and we have made a transformation to normal coordinates $\zeta(s) = N(s)z(s)$.

\paragraph{Birkhoff Normal Form}
For autonomous nonlinear Hamiltonian systems around an elliptic equilibrium, when we stay away from resonances, Birkhoff's theorem states that for a given polynomial order $N$ there exists a symplectic transformation such that the Hamiltonian in the new coordinates is broken into the normal form,
\[H^\prime = H(J) + R_{N+1},\]
where the first term depends only on the actions ($J_i = (q^2_i + p^2_i)/2$ in normal coordinates) and the second is a remainder term of polynomial order $N+1$ and up in the phase-space coordinates near the equilibrium.
In these normal coordinates, with small $R_{N+1}$, motion is again along circles in each phase-space plane, but with amplitude-dependent frequencies.
For systems with periodic time-dependence, the Poincaré map admits an analogous normal form which can be described through Lie algebraic techniques~\cite{dragtNormalFormMirror1979,forestNormalFormMethods1989,dragtLieAlgebraicTreatment1988}.
The time-dependent dynamics within one period of the system can be described through an $s$-dependent normalizing symplectic transformation and a phase that accumulates unevenly throughout time.
The evolution of the nonresonant nonlinear system which is well modeled by its series in Birkhoff normal form can be written as,
\[\mathcal{M}(s|s_0) = \mathcal{N}^{-1}(s)R(J,s,s_0)\mathcal{N}(s_0), \]
with $\mathcal{N}(s)$ as the time-dependent nonlinear normalizing transformation and $R(J,s, s_0)$ an amplitude- and time-dependent rotation~\cite{forestHamiltonianfreeDescriptionSingle1990,forestBeamDynamicsNew1998}.

\section{A Symplectic Flow in Normal Coordinates}
\label{sec:architecture}

We now turn to the problem of building conditioned generative models of physical systems.
Assume the samples $z$ in phase space are drawn from an unknown conditioned probability distribution $p_{\mathrm{data}}(z; c, c^*)$.
Here, we divide the conditioning parameters into two classes: those where the densities are related by symplectic transformation across their values ($c^*$, for example, a time-like coordinate), and those which are not ($c$, for example, values describing the initial state of some physical distribution evolving with time).
That is,
\[\rho(z; c, c^*_1) = \mathcal{M}(c^*_1|c^*_0)\rho(z; c, c^*_0),\]
such that  $\mathcal{M}(c^*_1|c^*_0)$ is symplectic.
We would like to find a model distribution that approximates $p_{\text{data}}$ well.

Which conditioning parameters belong in $c^*$ is determined by knowledge of the system.
The time-like coordinate is a trivial example of one of the symplectic conditioning parameters.
Due to the invertibility of classical mechanics, also included are system parameters when the densities for different values of the conditioning parameter are identical at some common time $s_0$ (for instance, time evolution from a common initial density).
We can time evolve from the first conditioning parameter value to $s_0$, change the conditioning parameter, which doesn't affect the density, and then evolve back to the time associated with the second value of the conditioning parameter; the composition is symplectic.

\begin{figure}
\centering
\includegraphics[]{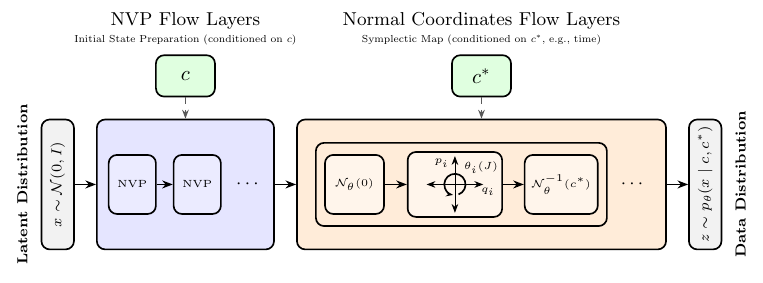}
\caption[]{Architecture of the conditioned normalizing flow showing evolution to an initial state, then transformation through the normal coordinates flow layers.}
\label{fig:normal-flow-architecture}
\end{figure}

The architecture proposed in this work is shown graphically in Fig.~\ref{fig:normal-flow-architecture}.
The normally distributed latent samples are first sent through typical normalizing flow layers, NVP flow~\cite{dinhDensityEstimationUsing2017} here, potentially conditioned on any non-symplectic parameters ($c$).
This allows the model to learn a family of distributions which the data distribution is related to by symplectic transformation (for example, preparing an initial state for time evolution).
That final mapping is accomplished with the normal coordinates flow conditioned on any symplectic parameters ($c^*$).
The initial NVP layers do not need to be symplectic and could be replaced with nearly any normalizing flow.
We chose Real NVP for its simplicity and because recent flows such as TarFlow~\cite{zhaiNormalizingFlowsAre2025} report their advantages on high-dimensional image data as opposed to the systems studied here.
These flows are not symplectic and cannot replace the normal coordinates flow layers themselves.
We note that the latent distribution can be chosen to be anisotropic, which may help downstream rotations be nondegenerate.

Inspired by the evolution of weakly nonlinear Hamiltonian systems in Birkhoff normal form, we seek an invertible symplectic flow layer consisting of a transformation to normal coordinates, an amplitude-dependent rotation, and a transformation out of normal coordinates.
That is,
\[ \mathcal{M} = \mathcal{N}^{-1}(c^*)R(J,c^*)\mathcal{N}(0),\]
where we have made the arbitrary choice of $c^*_0 = 0$ for the conditioning parameters corresponding to the initial distribution.
We drop $c^*_0$ from the rotation as we don't vary it in this model.
Each normal coordinates flow layer therefore involves one learned conditioned symplectic flow $\mathcal{N}(c^*)$ and a learned conditioned amplitude-dependent rotation flow $R(J,c^*)$, which may also depend on the conditioning parameters.
There are existing architectures for the general learnable symplectic transformation, and in this work we use SympNets (specifically LA-SympNets from Ref.~\cite{jinSympNetsIntrinsicStructurepreserving2020a}) initialized to identity transformations.
We will examine the rotational flow in the following subsection.

\subsection{Amplitude-Dependent Rotation Layer}

The rotational flow layer is a per-plane amplitude-dependent rotation,
\[R(J,c^*) = \begin{pmatrix} C(\theta(J,c^*)) & S(\theta(J,c^*)) \\ -S(\theta(J,c^*)) & C(\theta(J,c^*)) \end{pmatrix}, \qquad \begin{aligned} C(\theta) &= \mathrm{diag}\big(\cos\theta_1, \ldots, \cos\theta_n\big), \\ S(\theta) &= \mathrm{diag}\big(\sin\theta_1, \ldots, \sin\theta_n\big) \end{aligned}.\]
Transforming to action-angle coordinates via,
\[q_i = \sqrt{2 J_i}\cos\phi_i, \qquad p_i = \sqrt{2 J_i}\sin\phi_i,\]
the layer reduces to a translation of the angles by an action-dependent amount, leaving the actions invariant,
\[ \begin{pmatrix} J_i' \\ \phi_i' \end{pmatrix} = \begin{pmatrix} J_i \\ \phi_i + \theta_i(\boldsymbol{J}, c^\ast) \end{pmatrix}. \]
Noting that the transformation to action-angle coordinates is symplectic, our constraint that the rotation layer is symplectic requires that,
\[ \Omega^T\frac{\partial(\boldsymbol{J}', \boldsymbol{\phi}')}{\partial(\boldsymbol{J}, \boldsymbol{\phi})}\Omega = \Omega^T\begin{pmatrix} I & 0 \\ \partial \boldsymbol{\theta}/\partial \boldsymbol{J} & I \end{pmatrix}\Omega = \begin{pmatrix} \left(\partial \boldsymbol{\theta}/\partial \boldsymbol{J}\right) - \left(\partial \boldsymbol{\theta}/\partial \boldsymbol{J}\right)^T & I \\ -I & 0 \end{pmatrix} = \Omega,\]
which is only true when the Jacobian of $\theta(J,c^*)$ with respect to the actions is symmetric.
This is equivalent to the angles being derived from a scalar potential: $\theta(\boldsymbol{J},c^*) = \nabla_{\boldsymbol{J}} \Phi(\boldsymbol{J}, c^*)$.

To avoid the computational expense of calculating gradients with respect to the sample coordinates during training, we suggest using an architecture such as GradNets~\cite{chaudhariGradientNetworks2025}, which are constrained to learn functions derived from a scalar potential such as this.
In this work, we will use the cascaded gradient network ``GradNet-C'' architecture in particular and multiply the angles by a learnable logarithmic scaling factor initialized to $2\pi$.
Unfortunately, the network does not admit additional parameters that don't participate in the gradient, and so we require modification to allow it to depend on $c^*$.
We add an auxiliary conditioning multilayer perceptron which takes in the conditioning parameters $c^*$ and produces a latent conditioning vector $z$.
The conditioning vector is then read off with linear heads to derive the values $\alpha_i$, $\beta_i$, and $b_i$ from Ref.~\cite{chaudhariGradientNetworks2025} within each of the GradNet-C layers.

\section{Experiments}
\label{sec:experiments}

Here, we describe the results of three conditioned generative modeling experiments that were carried out in order to compare the normal coordinates flow layers to existing methods.
We have chosen one synthetic system and two physically motivated systems from the field of accelerator physics for this comparison.
The synthetic system is a chain of weakly interacting Duffing oscillators described by a quadratic Hamiltonian with a quartic nonlinearity whose strength we can vary.

In accelerator physics, charged particle beams are often modeled as a density in three physical dimensions (6D phase space) interacting with itself through the mean-field approximation.
Numerically, the beam is represented as samples from the phase space density, which makes building surrogate models of the outputs of codes simulating these systems a conditioned generative modeling problem.

The dynamics of the charged particle beam in the accelerator is described by a quadratic time dependent Hamiltonian with additional nonlinearities introduced through collective effects and nonlinear focusing elements.
This makes a normal coordinates flow a natural representation for time evolution in the system.
To this end, datasets were generated from simulation tools with the beam saved at evenly spaced time steps and the problem formulated as learning the beam distribution conditioned on the time-like coordinate.
For breadth, we have chosen two different collective effects commonly encountered in accelerator modeling for our two beam experiments: space charge and coherent synchrotron radiation.

Space charge describes the electrostatic repulsive force within the beam and, for most beam distributions, acts as a nonlinear time-dependent defocusing force.
Synchrotron radiation is the name for the light emitted by the beam when it is bent.
When the beam's length is shorter than the wavelength of the light it is producing, the process is coherent, with power scaling with the number of particles as $N^2$ instead of $N$ for the incoherent case.
The radiation field propagates along the chord of the bend in less time than the beam takes to traverse it and can affect the energy of particles (mostly $p_z$) in a time dependent and nonlinear way.
This shows up in transverse phase space ($x$ and $y$) due to the dispersion ($x$-$p_z$ and $y$-$p_z$ correlations) present in the bend.

For our baseline models, we have chosen Real NVP flows~\cite{dinhDensityEstimationUsing2017} and LA-SympNets~\cite{jinSympNetsIntrinsicStructurepreserving2020a}.
This is motivated by the use of SympNets in the flow layers themselves and by the need for a symplectic comparison flow.
For a fair comparison, the LA-SympNet architecture is allowed to have some number of unconditioned NVP flow layers after the latent and before the LA-SympNet modules act, just as for the fully integrated normal coordinates flow.
Additionally, in the linear blocks of the LA-SympNet, we use the more efficient parameterization of linear symplectic matrices described in Ref.~\cite{jinOptimalUnitTriangular2022}.
The Real NVP flow was conditioned by adding the conditioning parameters as input to the MLP in each block.
For the LA-SympNet, the parameters in each linear layer were taken as the output of an MLP that depends on the conditioning parameters.

All models were implemented in the Python programming language using PyTorch~\cite{paszkePyTorchImperativeStyle}.
For the experiments, hyperparameter tuning was carried out on the models using Bayesian optimization in Optuna~\cite{akibaOptunaNextgenerationHyperparameter2019}.
Each architecture within the experiments was allocated an equal amount of computational resources for tuning.
An NVIDIA RTX 4090 GPU was used for these experiments, with each model taking less than one hour to train and roughly three weeks of time in total for all tuning experiments.
The hyperparameters of the best models are listed in App.~\ref{sec:hyperparameters}.

\subsection{Duffing Chain}
For our first test, we chose to model the time evolution of a set of weakly interacting nonlinear oscillators.
A set of Duffing oscillators was augmented with a quadratic interaction term between chain-like neighbors.
The Hamiltonian for this system is,
\begin{equation}
	H(\mathbf{q}, \mathbf{p}) = \sum_{i=1}^{N} \left[ \frac{p_i^2}{2} + \frac{q_i^2}{2} + \frac{\beta}{4} q_i^4 + \frac{\epsilon}{2} \left( q_{i+1} - q_i \right)^2 \right], \qquad q_{N+1} \equiv q_1,
	\label{eq:duffing_chain}
\end{equation}
where $\beta$ is the strength of the nonlinearity, and $\epsilon$ is the interaction term.
In our experiment, the number of oscillators ($N$) was fixed to eight and the interaction term was $\epsilon=0.01$.
The time evolution of an ensemble of initial states was calculated numerically, and the distributions as a function of time were used in the generative modeling problem.
We integrated with velocity Verlet using a time step of $2\pi/1000$ and used the 152 snapshots, saved every 33 steps, up to $t=10\pi$.

The initial ensemble of 5k points was drawn from an uncorrelated Gaussian with standard deviations $(\sigma_{q_i}, \sigma_{p_i}) = (1.0, 0.25)$ for even $i$ and
$(0.25, 1.0)$ for odd $i$.
Isotropic distributions are invariant under rotation, requiring an elongated initial distribution to show the evolution and behavior like filamentation.
The distributions alternate so that neighboring oscillators are out of phase and may exchange energy through the coupling term.

\begin{figure}
	\centering
	\includegraphics{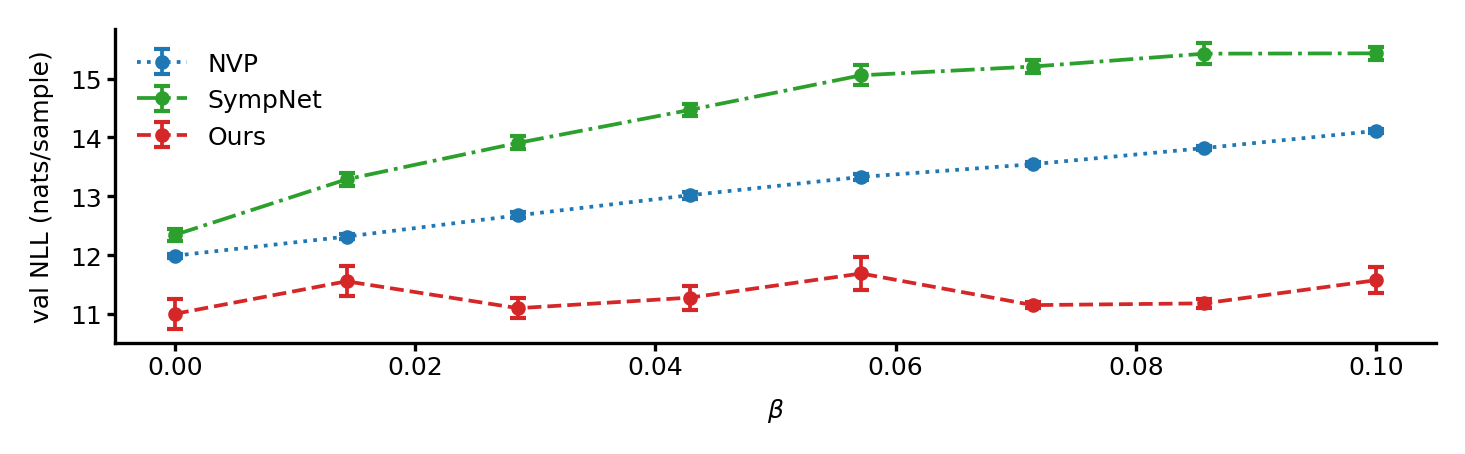}
	\caption[]{Held out negative log-likelihood on the Duffing chain example as the nonlinearity parameter ($\beta$) is swept. Points show mean and standard error.}
	\label{fig:duffing-beta-sweep}
\end{figure}

The synthetic dataset is convenient in that it allows us to vary the amount of nonlinearity in the system.
We performed experiments against the baseline models (Real NVP and LA-SympNet) as $\beta$ was swept from 0 to 0.1.
Hyperparameter tuning was performed once for each architecture for $\beta=0.1$.
Equal compute budget was allocated to each architecture and tuning was performed with the Optuna optimizer.
Architecture settings were varied along with learning rate and learning rate schedule (constant or cosine annealing with warm restarts~\cite{loshchilovSGDRStochasticGradient2017a}), with training fixed at roughly 1500 epochs.
The latent aspect ratio was also varied for the normal coordinates flow.
The best models were rerun 16 times at eight values of $\beta$ ranging from 0 to 0.1.
The mean and standard error of the negative log-likelihood (NLL) of the resulting models are shown in Fig.~\ref{fig:duffing-beta-sweep} as $\beta$ is changed.
Training was carried out with a 90\%-10\% train-test split such that samples from within a single snapshot do not show up in both the training and validation sets.
The split was stratified by quantile to avoid consecutive held out regions.

The normal coordinates flow already leads at $\beta=0$, the limit of linear dynamics.
As the nonlinearity parameter grows, the NLL of Real NVP and SympNet rises steadily from 12.0 to 14.1~nats/sample (Real NVP) and from 12.3 to 15.4~nats/sample (SympNet).
The normal coordinates flow remains consistent at 11--12~nats/sample across
the sweep.
This is supporting evidence for the architecture's inductive bias: amplitude-dependent rotations in normal coordinates.

\begin{figure}
	\centering
	\includegraphics{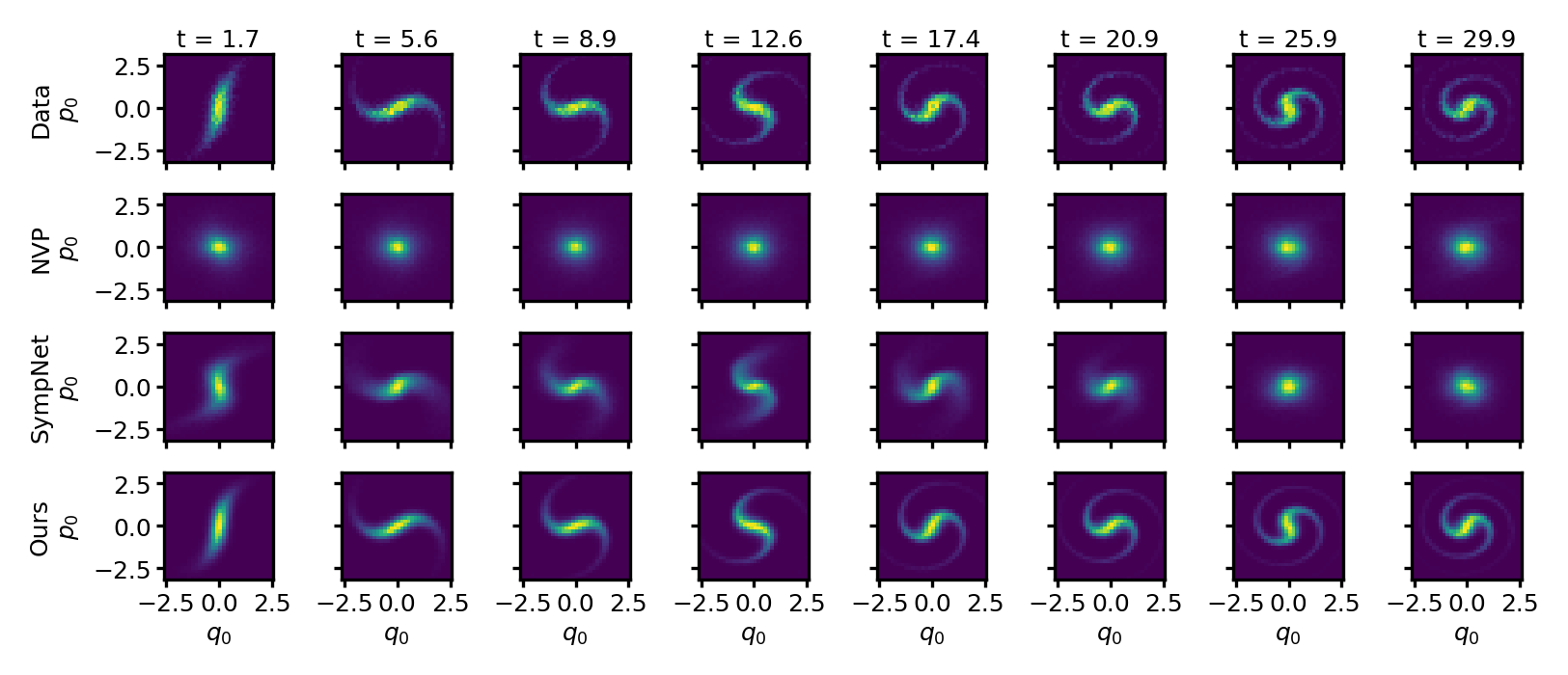}
	\caption[]{True and predicted phase space evolution of the ensemble of initial conditions for the first oscillator in the Duffing chain experiment with $\beta=0.1$.}
	\label{fig:duffing-beta-density}
\end{figure}

To better understand the behavior of the models on the nonlinear example, we plot samples from the models from the $\beta=0.1$ system against ground truth in Fig.~\ref{fig:duffing-beta-density}.
We show the ensemble's density in phase space for the first oscillator.
Here, it is clear that the normal coordinates flow is able to learn the dynamics of the system, including nonlinear features such as filamentation.
The SympNet model did learn a rotation of the initial Gaussian ensemble up until $t=12$, after which it deviated from ground truth.
The Real NVP model learned a degenerate Gaussian at the center of phase space without representing the rotation or nonlinear features.

Ablation studies performed on the Duffing chain system can be found in App.~\ref{sec:duffing-ablations}.

\subsection{Electron Beam with Space Charge in 1D Uniform Focusing Channel}

For a more physically motivated example, we perform tests on an electron beam in a uniform focusing channel with space charge.
Here, we model the beam with only one physical dimension (two dimensions in phase space).
Note that in 2D, preservation of the symplectic form is equivalent to preservation of volume, making all volume preserving transformations inherently symplectic.

In the uniform focusing channel, the beam feels a constant focusing force characterized by a strength $k=\qty{0.17}{\per\meter}$ such that with no other forces $x^{\prime\prime} = -k^2x$.
The amount of space charge force is characterized by the unitless beam perveance $K=4.2\times 10^{-5}$.
The initial beam is supergaussian distributed with supergaussian parameter $n=2.8$ (where 1 means a normal distribution), an emittance of $\varepsilon=\qty{0.65}{\micro\meter}$, and initial Twiss parameters $\beta_0=\qty{3}{\meter}$ and $\alpha_0=-0.34$.
The parameters in our system were chosen to produce interesting filamentation as the beam evolves under nonlinear space charge forces.

The system was simulated for \qty{300}{\meter} along the accelerator in a custom particle tracking code using 2M macroparticles.
There were 290 histograms of the 2D phase space saved at equally spaced locations along the lattice, from which samples were generated during model training by inversion sampling.
During training, a 90\%-10\% train-test split was used, with the split happening over the screens (i.e., each set of samples sharing a common $s$ position is either included or not included in one set; there is no leakage of samples within a value of the conditioning parameter).
The split was stratified by quantile to avoid consecutive held out regions.

\begin{figure}
	\centering
	\includegraphics{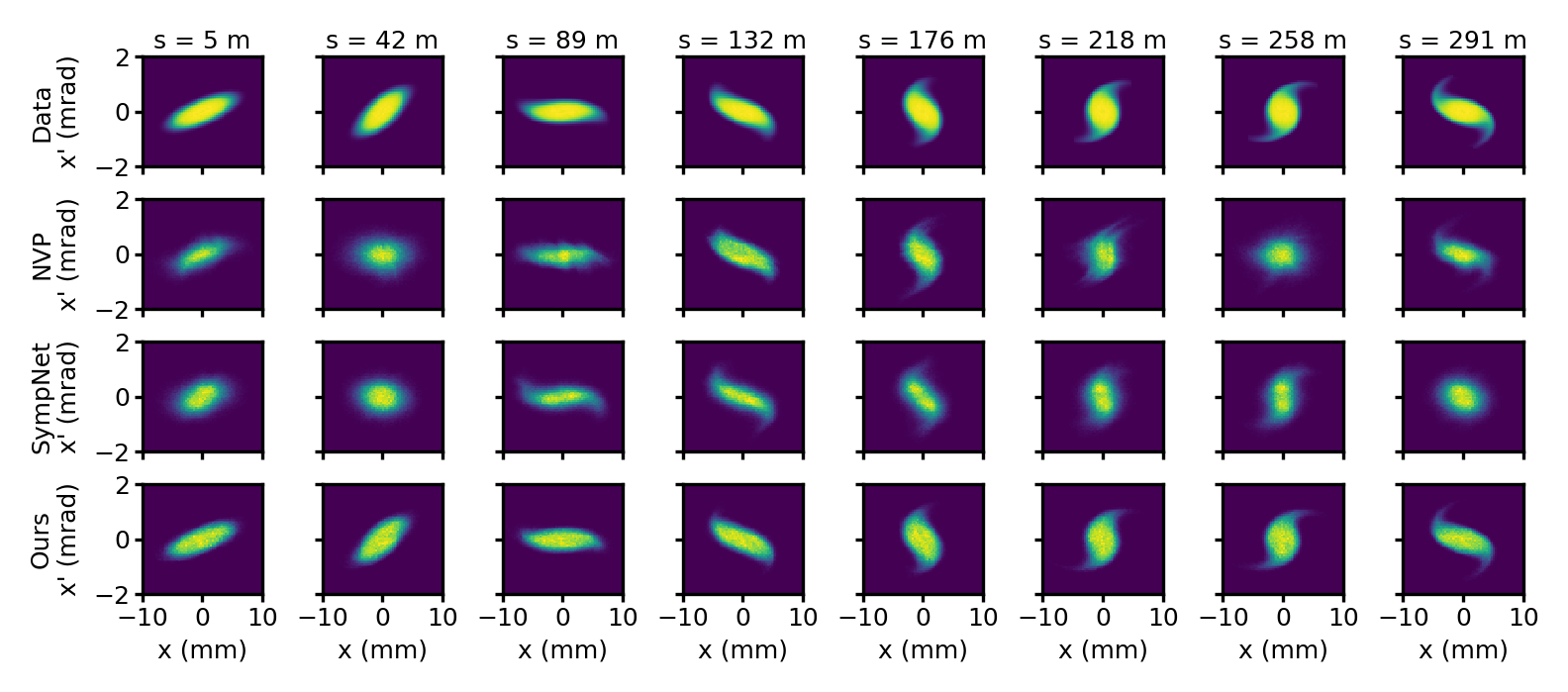}
	\caption[]{Models trained on the uniform focusing channel dataset. Histograms of samples drawn from the models at $s$ positions of held out test data compared with the ground-truth distribution.}
	\label{fig:focusing-channel-samples}
\end{figure}

Hyperparameter tuning was carried out on three model architectures: the normal coordinates flow, Real NVP flows, and LA-SympNets.
As mentioned earlier, an equal computational budget was dedicated to each architecture, and the tuning was carried out with Bayesian optimization in Optuna.
The best models were rerun five times with different seeds to develop statistics for the final negative log-likelihood validation loss.
The mean and standard error (1-$\sigma$, in parentheses) of the final loss were $-11.34$ (0.01) nats/sample for the Real NVP flow, $-11.25$ (0.03) nats/sample for the SympNet flow, and $-11.49$ (0.00) nats/sample for the normal coordinates flow.
Histograms of samples drawn from the best performing model in each architecture are shown in Fig.~\ref{fig:focusing-channel-samples} along with the ground-truth distribution.

\subsection{Electron Beam Through Compressor Chicane with Coherent Synchrotron Radiation}
\label{sec:chicane-experiment}

In our third experiment, we test the normal coordinates flows in three physical dimensions (6D phase space).
The data being modeled is generated from a simulation of an electron beam traversing a compressor chicane, as is used in accelerators such as free electron lasers.
A simple four-bend chicane with \qty{2.5}{\meter} legs and a \qty{5}{\meter} central section was designed to achieve \qty{10}{\milli\meter} of $R_{56}$ compaction (i.e., the linear $p_z\to t$ component in the transport map) with reasonable Twiss parameters throughout.
A schematic of the lattice and its associated optical functions is shown in Fig.~\ref{fig:chicane-layout} in App.~\ref{sec:chicane-additional-projections}.

The electron beam used in the experiment was \qty{1}{\nano\coulomb} at \qty{5}{\giga\electronvolt} and began with a normalized emittance of \qty{10}{\micro\meter}.
In the $x$/$y$ planes, the beam was supergaussian with the parameter $n=4$ ($n=1$ means normally distributed).
Longitudinally, the beam followed a Tukey distribution with a full length of \qty{30}{\micro\meter} and a taper ratio of 0.75, and the momentum distributions were all Gaussian.
The tool Distgen~\cite{gullifordDistgen2026} was used to generate the beam with 100k macroparticles.
Additional nonlinearities were then added to the beam along with a chirp of $R_{56} = \qty{-25}{\milli\meter}$ to make for an interesting, nontrivial initial distribution.
This was then sent to Bmad~\cite{saganBmadRelativisticCharged2006} for particle tracking with coherent synchrotron radiation (CSR) calculation included.
The beam was found to lose \qty{7}{\mega\electronvolt} (0.1\%) of energy to CSR through the chicane.
We note that lack of energy conservation means the time evolution of the system is not strictly symplectic.
However, for small energy losses, it will remain a good approximation.

From the simulation tool, the set of all particle coordinates in 6D phase space was saved at 128 evenly spaced points along the lattice.
It is this data which forms the basis of our conditional generative modeling problem: learn the phase space densities as a function of position along the lattice.

As in the earlier experiments, a 90\%-10\%  stratified train-test split was carried out with each output at a single position being either included in one set or not to avoid leakage.
Our initial tests showed that with enough parameters, all models can learn the conditioned data equally well (achieving roughly $-5.2$ nats/sample negative log-likelihood).
To create a test capable of distinguishing the models on how efficiently they are able to learn the data instead, we restrict ourselves to a comparison of all three model architectures with at most 20k parameters and 3k epochs for training.
All models were tuned with equal computational budgets using Optuna Bayesian optimization.

\begin{figure}
	\includegraphics{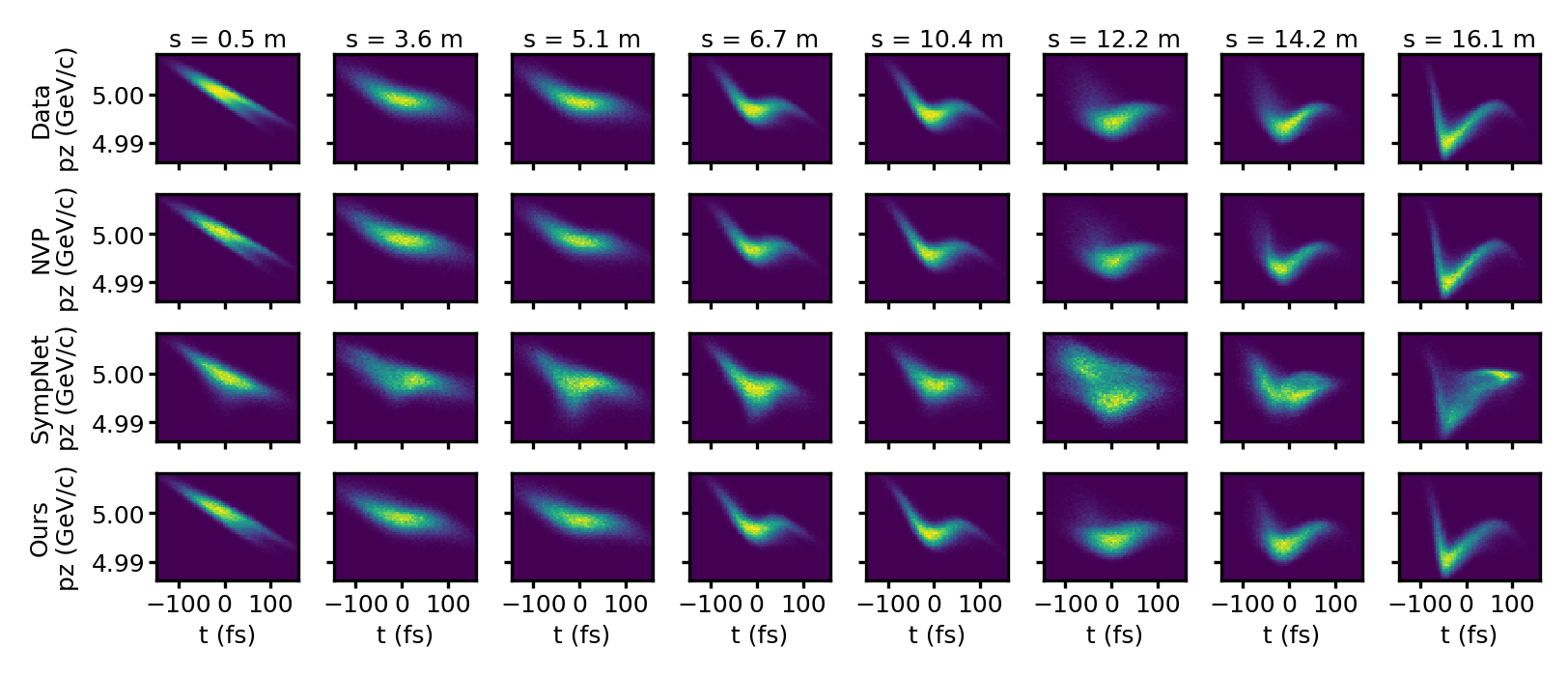}
	\caption[]{$t$-$p_z$ beam distributions from the held out test data in the chicane problem. (top) ground-truth distribution; (top-middle) distribution from the best Real NVP flow; (bottom-middle) distribution from the best SympNet flow; (bottom) distribution from the normal coordinates flow.}
	\label{fig:chicane-t-pz}
\end{figure}

\begin{figure}
	\includegraphics{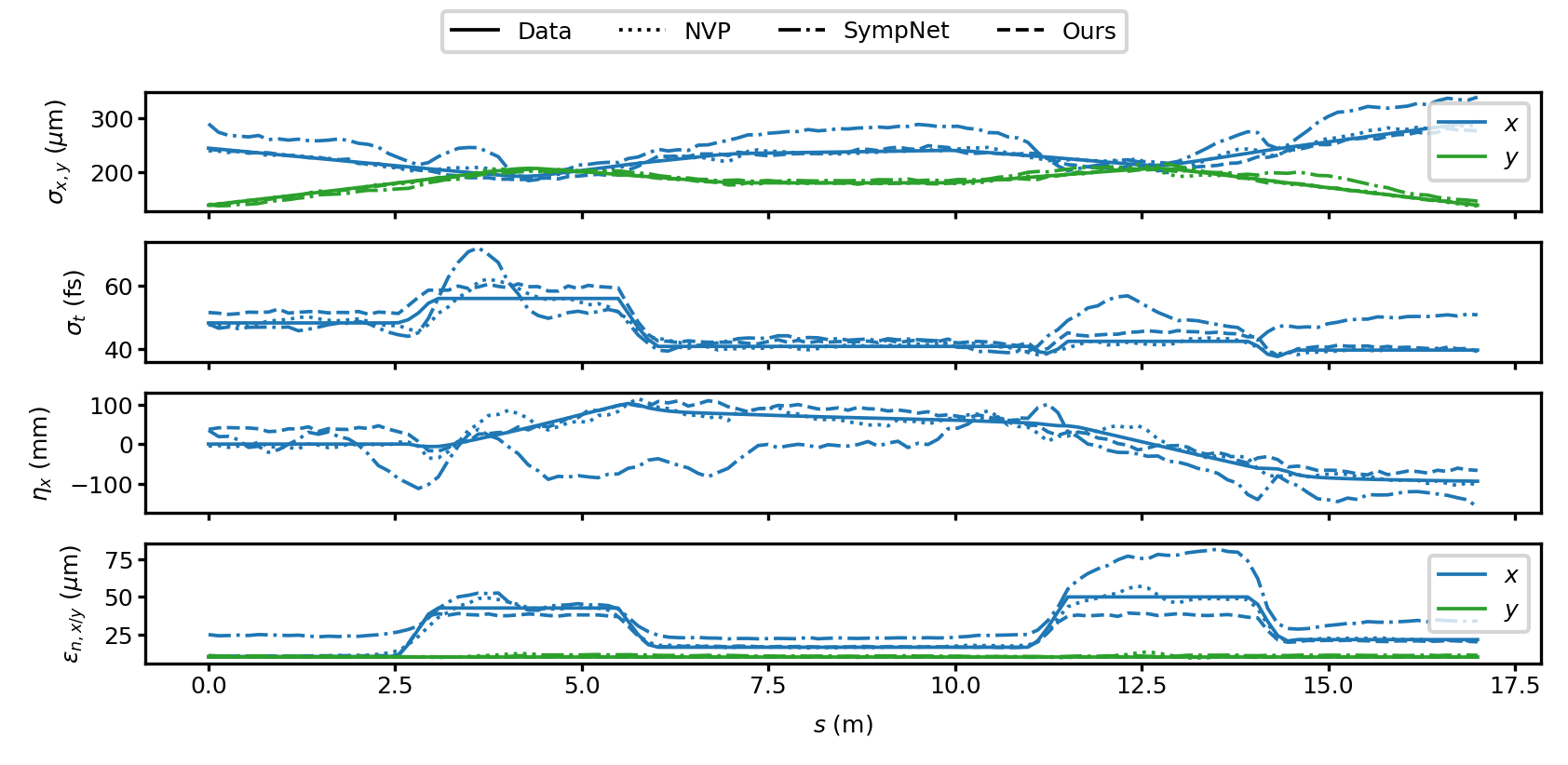}
	\caption[]{Evolution of the beam's second moments along the chicane lattice. Beam size ($\sigma_{x/y}$), bunch length ($\sigma_t$), dispersion ($\eta_x$), and normalized emittance ($\varepsilon_{n,x/y}$) are all shown. The cosine similarity of each model to the ground truth is given in Tab.~\ref{tab:chicane-similarity}.}
	\label{fig:chicane-rms}
\end{figure}

The best performing models of each architecture were rerun five times with different seeds.
The mean and standard error of the final negative log-likelihood on validation samples (1-$\sigma$, in parentheses) were $-5.06$ (0.01) nats/sample for the Real NVP model, $-4.15$ (0.03) nats/sample for the SympNet model, and $-5.07$ (0.02) nats/sample for the normal coordinates flow.
Due to space constraints, we can only show one of the three phase space plane projections in this article, with the rest reserved for App.~\ref{sec:chicane-additional-projections}.
We choose the $t$-$p_z$ plane as it includes the most interesting nonlinearities in the system.
In Fig.~\ref{fig:chicane-t-pz}, histograms showing the evolution of the beam in longitudinal phase space are shown along with outputs from each model.
We also show some of the second moments of the beam distribution and the comparisons with the models in Fig.~\ref{fig:chicane-rms}.
The mean-subtracted cosine similarity between each model's trace and the ground truth data is shown in Tab.~\ref{tab:chicane-similarity} in App.~\ref{sec:chicane-additional-projections}.

\section{Conclusion}
\label{sec:conclusion}

In this work, we have explored a new normal coordinates flow layer inspired by the dynamics of nonlinear Hamiltonian systems in Birkhoff normal form.
Here, after a symplectic transformation, the dynamics is reduced to amplitude-dependent rotations.
We have shown that it is possible to construct a nonlinear rotation layer for the flow and that, as long as the Jacobian of the angles with respect to the actions is symmetric, the transformation is symplectic.
This condition is met by a GradNet.
The new normal coordinates flow layer was then tested on three conditioned generative modeling problems for physical systems.
On the Duffing chain, it outperformed both baseline models across the nonlinearity sweep by a margin that grew with the strength of the nonlinearity.
Ablation studies on this system showed that amplitude dependence and the rotation layer are both critical to this performance.
The flow showed superior results to both baseline models in the case of the uniform focusing channel.
For the chicane problem, it was more efficient than the SympNet and showed equal performance to Real NVP flows.

A key limitation of the architecture is the restriction of the conditioned transformation to rotations in normal coordinates.
We take inspiration from Birkhoff's normal form, but this argument is not true when resonance driving terms are present.
Additionally, Birkhoff's normal form is a formal series expansion of the Hamiltonian and may diverge.
In this case, the remainder term stays large in the region of phase space we want to model, and the dynamics may not simply reduce to amplitude-dependent rotations.
Finally, our experiments are limited to simulated systems of up to 16 phase-space dimensions.

One future line of work is exploring symplectic flow layers more expressive than the LA-SympNets used here to construct the transformation to normal coordinates.
The architecture described in this work benefits from any advancements in symplectic flow layers, as they can immediately be integrated with the rotational layers to form a normal coordinates flow.
It would be an interesting experiment to see which layers are most commensurate with learning the transformation to normal coordinates.
The initial NVP layers could similarly be replaced with more recent normalizing flows.

The model described in this work enforces hard symplectic constraints through architecture.
Additionally, amplitude-dependent rotations in phase space are a natural way to express the dynamics of weakly nonlinear Hamiltonian systems.
Normal coordinates flows should provide an inductive bias towards models that represent these systems effectively.
These results add to the practitioner's growing toolbox of architectures tailored for learning the typical dynamics of physical systems.

\begin{ack}
The author has no funding or competing interests to declare.
\end{ack}

\printbibliography[title={References}]

\clearpage
\appendix
\section{Additional Results from Chicane Experiment}
\label{sec:chicane-additional-projections}
\begin{figure}
	\includegraphics{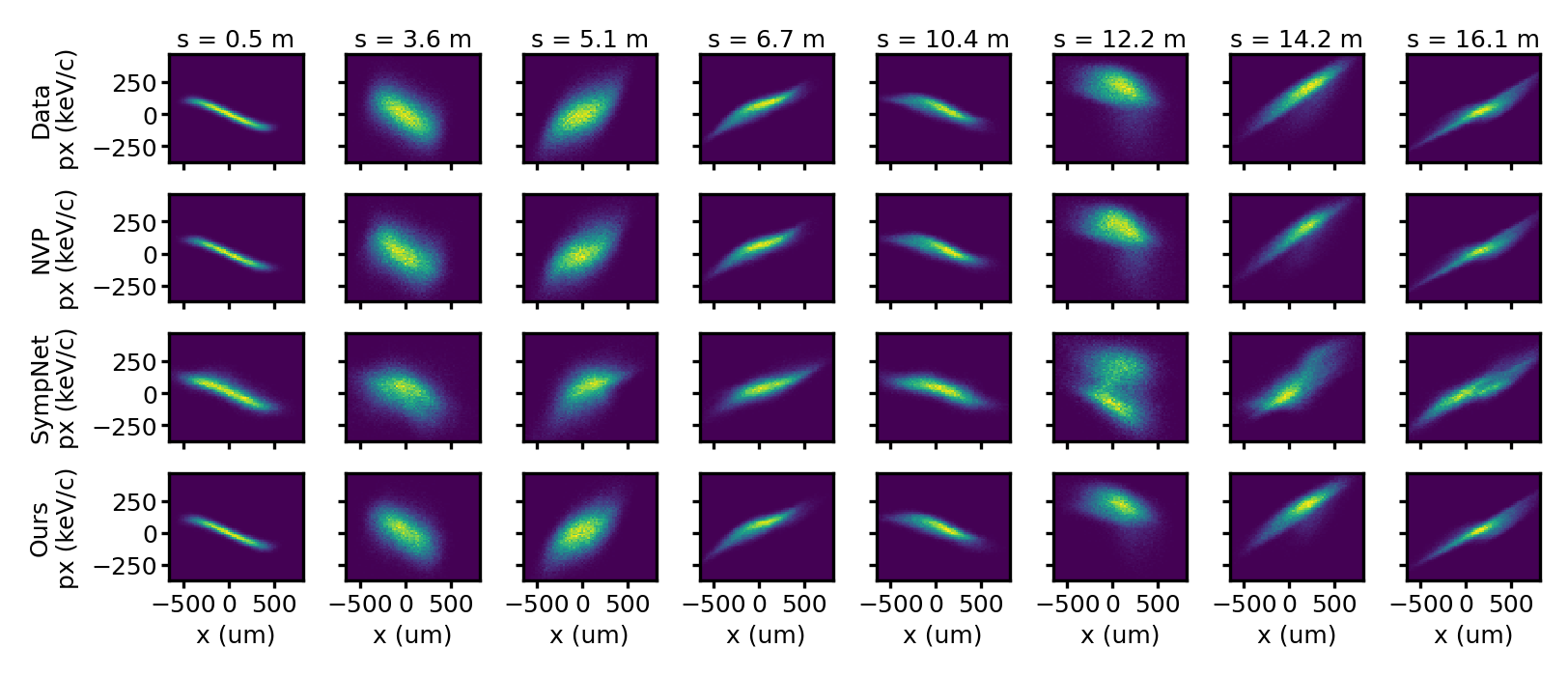}
	\caption[]{Held out examples from the chicane dataset showing the $x$-$p_x$ phase space projection. Same model labels as in Fig.~\ref{fig:chicane-t-pz}.}
	\label{fig:chicane-x-px}
\end{figure}

\begin{figure}
	\includegraphics{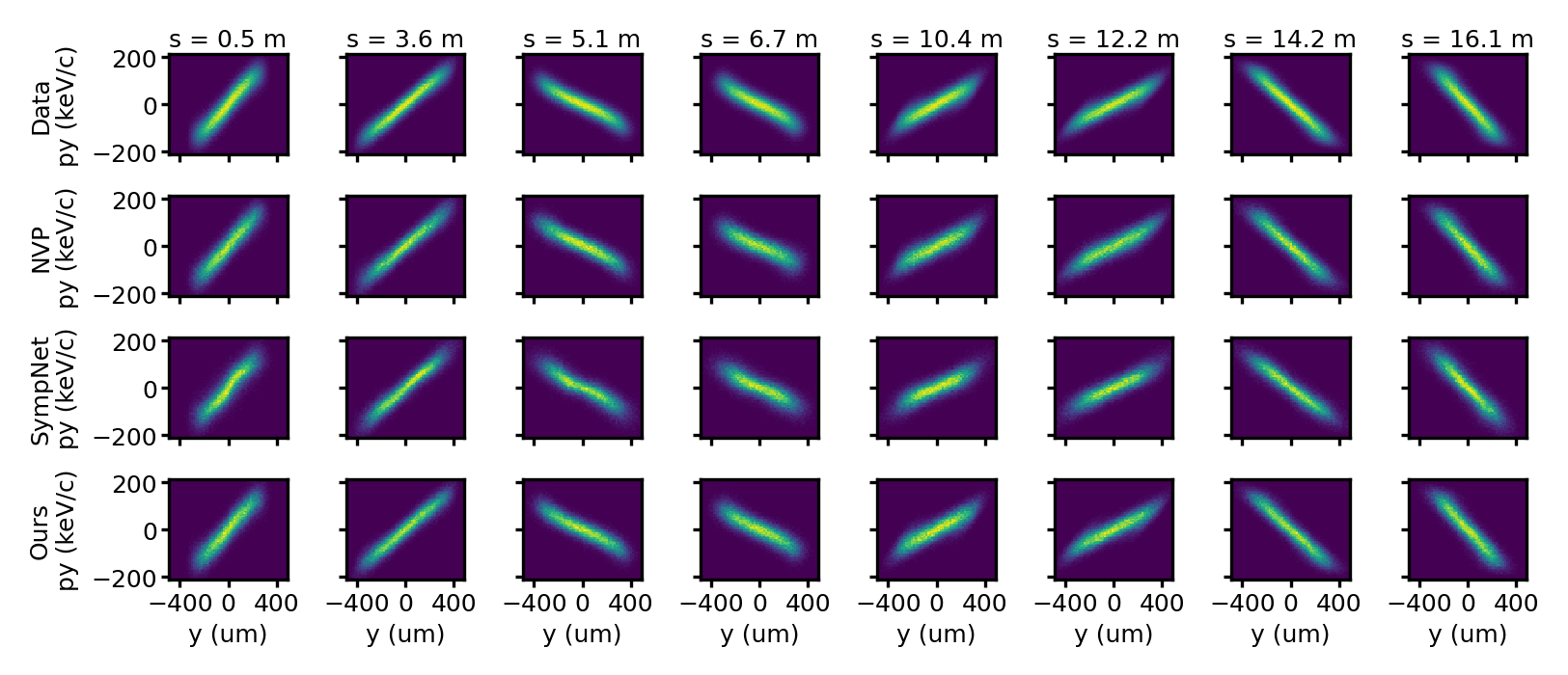}
	\caption[]{Held out examples from the chicane dataset showing the $y$-$p_y$ phase space projection. Same model labels as in Fig.~\ref{fig:chicane-t-pz}.}
	\label{fig:chicane-y-py}
\end{figure}

In this brief appendix, we include the lattice schematic, the cosine similarity of the beam statistics, and the remaining phase space projections from Sec.~\ref{sec:chicane-experiment} due to lack of space in the main body of the article.
The two transverse phase spaces are shown in Fig.~\ref{fig:chicane-x-px}, for $x$-$p_x$, and Fig.~\ref{fig:chicane-y-py}, for $y$-$p_y$.
All samples are drawn from the same models as in Fig.~\ref{fig:chicane-t-pz}.

\begin{figure}[h]
	\centering
	\includegraphics{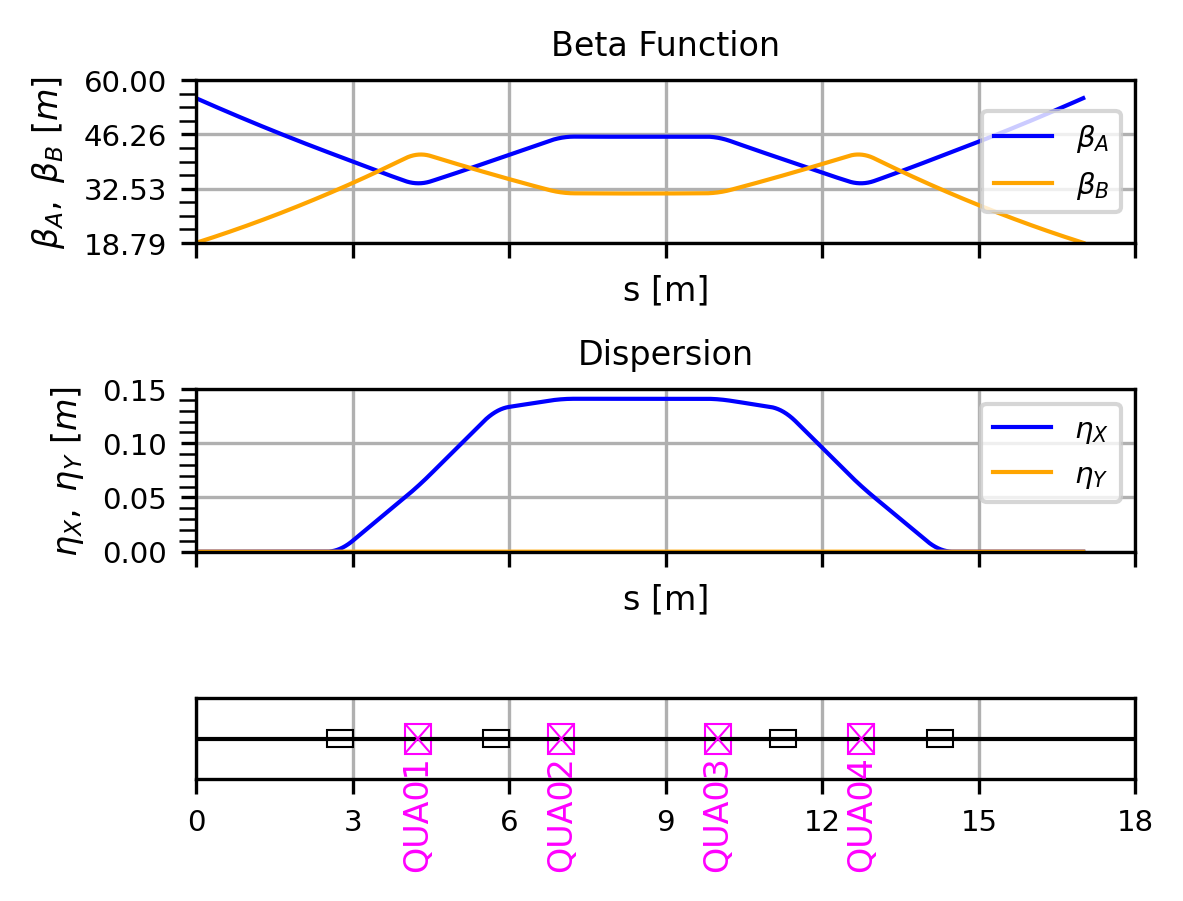}
	\caption{Layout of the chicane used in the 6D generative modeling experiment. (top) beta functions showing the beam envelope throughout the lattice; (middle) dispersion, showing the $x$-$p_z$ correlation of the beam; (bottom) the layout with dipole (bending) magnets in black and quadrupole (focusing) magnets in magenta.}
	\label{fig:chicane-layout}
\end{figure}

\begin{table}
	\caption{
		Mean-subtracted cosine similarity between model and ground-truth traces from Fig.~\ref{fig:chicane-rms} over all 128 lattice positions.
		The ground truth vertical emittance ($\varepsilon_{n,y}$) does not change across $s$, causing cosine similarity to only show the residual noise in the models.
		Because of this, we don't include it in this table.
	}
	\label{tab:chicane-similarity}
	\centering
	\begin{tabular}{lcccccc}
		\toprule
		Model & $\sigma_x$ & $\sigma_y$ & $\sigma_t$ & $\eta_x$ & $\varepsilon_{n,x}$ & Average \\
		\midrule
		Real NVP & 0.974 & 0.988 & 0.960 & 0.961 & 0.987 & 0.974 \\
		SympNet & 0.913 & 0.924 & 0.589 & 0.610 & 0.915 & 0.790 \\
		Normal coordinates & 0.969 & 0.997 & 0.994 & 0.960 & 0.986 & 0.981 \\
		\bottomrule
	\end{tabular}
\end{table}

\clearpage
\section{Normal Coordinates Flow Ablation Studies}
\label{sec:duffing-ablations}
In this section, we describe the results of ablation studies performed on the normal coordinates flow architecture.
The purpose of these studies is to provide experimental evidence for the necessity of parts of the architecture to its performance.
All experiments were performed on the Duffing chain dataset with $\beta=0.1$ as it provides a challenging example for generative modeling (as seen from the performance of the SympNet and Real NVP models on it).
Hyperparameters were the best found in our tuning experiment (App.~\ref{sec:hyperparameters}) except where noted otherwise.

In the first experiment, we ablated each important characteristic of the normal coordinates flow architecture.
Amplitude dependence was removed from the rotation layer by zeroing out the actions.
The whole rotation layer was removed from the architecture so that only the transformations to and from normal coordinates remain.
The rotation layer was made non-symplectic by replacing the GradNet with an MLP of similar capacity.

Each experiment was repeated at least three times for statistics (except for the baseline, where all 16 values from the sweep run were used) and the results are shown in Tab.~\ref{tab:duffing-ablations}.
The baseline from the Duffing chain section is also shown.
Each of the ablations degraded the performance of the network, with particularly severe results for removal of amplitude dependence and removal of the rotation layer.
Removing symplecticity of the rotation layer still degrades performance by 1.4~nats/sample (1.6 standard deviations above baseline).

The sensitivity of the architecture to the conditioning network's capacity was also investigated.
The layer width was swept over 16, 32, 64, and 128 units with other hyperparameters held the same.
Models were trained at least three times each (again, except for the baseline, which had 16 values) for statistics, and the resulting NLL is documented in Tab.~\ref{tab:duffing-cond-width}.
The likelihoods stay within one standard deviation of the baseline value as the network is varied.

These results suggest that amplitude-dependent symplectic rotation is important in modeling systems similar to the Duffing chain.
This contrasts the normal coordinates architecture with existing work related to normal coordinate maps~\cite{duruisseauxApproximationNearlyperiodicSymplectic2023} that does not include amplitude dependence.

\begin{table}[h]
	\caption{
		Ablation studies of the normal coordinates flow on the Duffing chain with $\beta=0.1$. Mean and standard deviation (in parentheses) of the validation NLL are shown.
	}
	\label{tab:duffing-ablations}
	\centering
	\begin{tabular}{lc}
		\toprule
		Experiment & NLL (nats/sample) \\
		\midrule
		Baseline & 11.6 (0.9) \\
		Remove amplitude dependence of rotation layer & 17.0 (0.4) \\
		Remove rotation layer & 16.7 (0.1) \\
		Use MLP instead of GradNet in rotation layer & 13.0 (0.5) \\
		\bottomrule
	\end{tabular}
\end{table}

\begin{table}[h]
	\caption{
		Sensitivity to the conditioning network layer width on the Duffing chain with $\beta=0.1$.
		Mean and standard deviation (in parentheses) of the validation NLL are shown.
	}
	\label{tab:duffing-cond-width}
	\centering
	\begin{tabular}{lc}
		\toprule
		Conditioning Network Layer Width & NLL (nats/sample) \\
		\midrule
		16 & 11.2 (0.3) \\
		32 & 11.6 (0.8) \\
		64 (baseline) & 11.6 (0.9) \\
		128 & 11.2 (0.2) \\
		\bottomrule
	\end{tabular}
\end{table}

\section{Hyperparameters of the Best Models}
\label{sec:hyperparameters}

\paragraph{Duffing Chain}
The leaky ReLU activation function was used for coupling layer MLPs, with rotation networks using GELU.
All models were trained with the Adam optimizer~\cite{kingma_adam_2017} on negative log-likelihood loss.
The Real NVP flow used 10 coupling layers with 2 hidden layers of 128 units each, with a learning rate of $3.0\times10^{-3}$ for 1500 epochs and a batch size of 32.
The LA-SympNet used 4 initial unconditioned NVP layers (MLPs with 4 hidden layers of 16 units) followed by 10 LA-SympNet modules, trained with cosine annealing with warm restarts ($T_0 = 100$ epochs, 3 restarts) with peak learning rate $2.8\times10^{-3}$ and minimum $2.8\times10^{-5}$ and batch size 32.
The normal coordinates flow used four initial unconditioned NVP layers (1 hidden layer, 64 units) followed by four normal coordinates flow layers.
Each used two LA-SympNet modules for the normalizing transformation, a rotation layer with two hidden layers of 64 units, and a conditioning network using two hidden layers of 64 units and a latent conditioning dimension of eight.
It was trained with cosine annealing with warm restarts ($T_0 = 214$ epochs, 2 restarts) with peak learning rate $2.5\times10^{-3}$ and minimum $2.5\times10^{-5}$ and batch size 32.
Its latent base distribution was anisotropic with $\sigma_p/\sigma_q = 3.8$ in each phase-space plane.

\paragraph{Uniform Focusing Channel}
All models used leaky ReLU for the MLP activation functions, with rotation networks using GELU, and were trained with the Adam optimizer~\cite{kingma_adam_2017} on negative log-likelihood loss.
Models were given several learning rate schedules to choose from using cosine annealing with warm restarts~\cite{loshchilovSGDRStochasticGradient2017a}.
The best performing Real NVP flow used 12 coupling layers with MLPs of 4 hidden layers of 64 units each, trained at a constant learning rate of $1.1\times10^{-3}$ for 750 epochs with batch size 8.
The best LA-SympNet used 4 initial unconditioned NVP layers (MLPs with 3 hidden layers of 32 units) followed by 10 LA-SympNet modules, trained at a constant learning rate of $2.5\times10^{-4}$ for 750 epochs with batch size 8.
The normal coordinates flow with the best performance had 4 initial unconditioned NVP layers (MLPs with 3 hidden layers of 32 units) followed by 4 normal coordinates flow layers, each with 3 LA-SympNet modules for the normalizing transform.
The rotation networks used 4 hidden layers of 32 units with a conditioning network of 4 hidden layers of 64 units and a latent conditioning dimension of 8.
It was trained with cosine annealing with warm restarts ($T_0 = 50$ epochs, 3 restarts) with peak learning rate $2.5\times10^{-4}$ and minimum $10^{-6}$ and batch size 16.

\paragraph{Compressor Chicane}
We again used leaky ReLU for the MLP activation functions, with rotation networks using GELU, and models were trained with the Adam~\cite{kingma_adam_2017} optimizer on negative log-likelihood loss with the option to use cosine annealing with warm restarts~\cite{loshchilovSGDRStochasticGradient2017a}.
The best performing Real NVP flow used 10 coupling layers with MLPs of 2 hidden layers of 32 units each.
It was trained with cosine annealing with warm restarts ($T_0 \approx 50$ epochs, 5 restarts) with peak learning rate $2.4\times10^{-3}$ and minimum $2.4\times10^{-5}$ and batch size 16.
The best LA-SympNet used 4 initial unconditioned NVP layers (MLPs with 1 hidden layer of 32 units) followed by 10 LA-SympNet modules, trained at a constant learning rate of $1.2\times10^{-3}$ for 3050 epochs with batch size 16.
The normal coordinates flow with the best performance had 4 initial unconditioned NVP layers (MLPs with 1 hidden layer of 16 units) followed by 1 normal coordinates flow layer with 8 LA-SympNet modules for the normalizing transform.
The rotation networks used 3 hidden layers of 32 units with a conditioning network of 3 hidden layers of 16 units and a latent conditioning dimension of 8.
It was trained with cosine annealing with warm restarts ($T_0 \approx 25$ epochs, 6 restarts) with peak learning rate $9.8\times10^{-3}$ and minimum $1.1\times10^{-5}$ and batch size 16.

\end{document}